\documentclass[aps,prl,twocolumn,showpacs,superscriptaddress,amsmath,amssymb]{revtex4}

\usepackage{graphicx}

\begin{document}



\title{A Stochastic Feedback Model for Volatility}



\author{Raoul Golan} \affiliation{Finance Discipline Group, University of Technology, Sydney, Broadway, NSW 2007, Australia}

\author{Austin Gerig} \email{austin.gerig@sbs.ox.ac.uk}
\affiliation{CABDyN Complexity Centre, Sa\"{i}d Business School, University of Oxford, Oxford OX1 1HP, United Kingdom}



\begin{abstract}

Financial time series exhibit a number of interesting properties that are difficult to explain with simple models.  These properties include fat-tails in the distribution of price fluctuations (or \emph{returns}) that are slowly removed at longer timescales, strong autocorrelations in absolute returns but zero autocorrelation in returns themselves, and multifractal scaling. Although the underlying cause of these features is unknown, there is growing evidence they originate in the behavior of \emph{volatility}, i.e., in the behavior of the magnitude of price fluctuations.  In this paper, we posit a feedback mechanism for volatility that closely reproduces the non-trivial properties of empirical prices.  The model is parsimonious, contains only two parameters that are easily estimated, fits empirical data better than standard models, and can be grounded in a straightforward framework where volatility fluctuations are driven by the estimation error of an exogenous Poisson rate.

\end{abstract}



\maketitle



Financial time series exhibit a number of interesting regularities (or ``stylized facts'', as economists call them) that are well-documented in the literature\cite{Mandelbrot63, Fama65, Bollerslev1994, Guillaume1997, Cont2001, BouchaudPotters03, Borland2005}. They include fat tails in the distribution of price fluctuations (known as \emph{returns} in finance) that are slowly removed at longer timescales, strong autocorrelations in absolute returns but zero autocorrelation in returns themselves, multifractal scaling, and a negative correlation between past returns and the future magnitude of returns (known as the leverage effect). 

Although there is currently no consensus on the underlying cause of these properties, there is growing evidence they are all rooted in the behavior of \emph{volatility}, i.e., in the behavior of the magnitude of returns\cite{Bollerslev1994}.  In addition, there is evidence that returns -- like earthquakes, turbulent flow, and Barkhausen noise -- are driven by strong endogenous, or internal, feedback effects\cite{Bouchaud11}.

In this paper, we present a stochastic feedback model for volatility that generates many of the stylized facts of financial time series.  The model is motivated by several recent studies that have found the variance of price fluctuations, i.e., the squared volatility, is slowly varying and inverse gamma distributed\cite{Gerig2009,Fuentes2009,Ma2013} so that returns are well-fit by a Student's $t$-distribution at short timescales\cite{Gerig2009,Fuentes2009,Ma2013,Praetz72,Blattberg74,Peiro1994,Tsallis03,Platen2008,Gu08,Gerig2011}.  Here we extend these results by modelling the properties of returns over timescales longer than one day.  We propose a simple mechanism that generates inverse gamma distributed variances and introduce a feedback parameter that allows the variance to change slowly over time as it does in real price series.  As a result, returns are $t$-distributed at daily intervals but slowly approach a Gaussian as timescales are increased to weekly, monthly, and yearly intervals. 

Although the results of the model match empirical data very well, we make no strong claim that we have uncovered \emph{the} mechanism driving real-world volatility fluctuations.  Instead, we offer the proposed mechanism as a novel explanation for these fluctuations and leave any conclusions regarding the true mechanism for later analysis. Alternative models that also produce inverse gamma distributed variances (and therefore Student $t$-distributed returns) include the well-known GARCH model\cite{Bollerslev1986, Nelson1990, Bollerslev1994} which can be motivated by the position of stop-loss orders in markets\cite{Borland2011} and the Minimal Market Model\cite{Platen2001,PlatenHeath2006} which describes the dynamics of a growth optimal portfolio with deterministic drift.

To begin our analysis, we define the $t^{th}$ return as the difference in logarithmic price from time $t$ to time $t+\Delta t$ where $t$ is measured in days,
\begin{equation}
r_t(\Delta t) = \ln{(p_{t+\Delta t})} - \ln{(p_t)}.
\end{equation}
We model daily returns, $r_t(\Delta t=1)$, as a discrete time stochastic process with a fluctuating variance, 
\begin{equation}
r_t = \mu + \sigma_t \xi_t,
\label{eq.arch}
\end{equation}
where $\mu$ is the daily average return, $\xi_t$ is an IID Gaussian $N(0,1)$ random variable, and $\sigma_t^2$ is the local variance of returns (the square of the daily volatility, $\sigma_t$).  

We assume that the daily variance of returns, $\sigma_t^2$, is determined by a stochastic feedback process so that, 
\begin{equation}
\sigma^2_{t} \sim \frac{\sigma^2_0}{Gamma(1+B\sigma^2_0/\sigma^2_{t-1},1+B)},
\end{equation}
where $Gamma(a,b)$ is the gamma distribution, $f(x|a,b)=(b^a/\Gamma[a])x^{a-1}e^{-bx}$.  

Only two parameters are used in Eq.~3: $1/\sigma^2_0$ is the equilibrium inverse variance of the return process and $B>1$ is a feedback parameter.  Notice that when $\sigma^2_{t-1}=\sigma^2_0$, the expected value of $1/\sigma^2_{t}$ is $1/\sigma^2_0$.  Deviations from this equilibrium value are removed over a length of time determined by $B$.  If $B$ is large, the variance requires many iterations to relax back towards $\sigma^2_0$, but for small $B$, the relaxation is quick.

The model can be motivated by the following mechanism: suppose that market participants continually observe an exogenous Poisson process that influences how they trade (this process could describe, for example, the arrival of new orders to the market or the arrival of economic news).  When they estimate a high rate for the process, they increase their activity in the market and raise market volatility.  Likewise, when they estimate a low value for this rate, they decrease their activity and reduce market volatility.  

Specifically, assume that detrended intraday returns are uncorrelated, are all of size $\pm\delta$, and occur at an average daily rate, $\lambda(t)$, that changes from day to day.  $\lambda(t)$ is influenced by market participants as follows: (1) they observe an exogenous Poisson process with rate parameter $\lambda_e$, (2) based on the estimated rate of this process, they act in the market such that, on average, $N$ price fluctuations occur per $M$ exogenous events.  Denoting their estimate of the exogenous rate by $\widehat{\lambda}_e(t)$, 
\begin{equation}
\lambda(t) = (N/M) \widehat{\lambda}_e(t).
\end{equation}
$M$ exogenous events will occur in an amount of time, $\tau\sim Gamma(M,\lambda_e)$, making the estimated rate inverse gamma distributed, $\widehat{\lambda}_e(t)\sim M/Gamma(M,\lambda_e)$.  The local variance of daily returns is therefore,
\begin{equation}
\sigma_t^2 = \delta^2 \lambda(t) \sim \frac{\delta^2 N }{Gamma(M,\lambda_e)}.
\end{equation}
Notice that this mechanism produces inverse gamma distributed variances, but these variances are not autocorrelated.

To introduce feedback effects we assume that $M$ varies through time so that the sensitivity of the market to exogenous events (measured by $N/M$) is high when the local variance is high and low when the local variance is low.  The simplest form of this relationship (taking into account that $M\ge1$) is,
\begin{equation}
M(t) = 1 + A/\sigma_{t-1}^2.
\end{equation}
The final result is,
\begin{equation}
\sigma_t^2 \sim \frac{\delta^2 N }{Gamma(1 + A/\sigma_{t-1}^2,\lambda_e)}.
\end{equation}
The equation can be simplified by introducing the equilibrium inverse variance $1/\sigma^2_0$, defined by the relation $E[1/\sigma_t^2|\sigma_{t-1}^2 = \sigma_0^2] = 1/\sigma_0^2$.  Therefore, $\sigma_0^2 = \delta^2 N \lambda_e - A$.  Using this relation and the simplifying assumption that at equilibrium, $E[\tau|\sigma_{t-1}^2 = \sigma_0^2]=1$, so that $A = \sigma_0^2(\lambda_e - 1)$, we have, 
\begin{equation}
\sigma_t^2 \sim \frac{\sigma_0^2 }{Gamma(1 + (\lambda_e-1)\sigma_0^2/\sigma_{t-1}^2,\lambda_e)}.
\end{equation}
Notice this equation is identical to Eq.~3 with the change of variable $\lambda_e=B+1$.  According to the proposed mechanism, $B+1$ is the daily rate of the exogenous process.  If $B$ is large, the exogenous process has a high daily rate and can be estimated without too much error by market participants.  Volatility therefore does not change much day to day and is strongly autocorrelated.  If $B$ is small, the exogenous process has a low rate and there are large errors in its estimation, which means volatility autocorrelations are less strong.

To determine how the parameters of the model affect returns, we simulate the model using different choices of $\sigma_0^2$ and $B$ (each run includes 50 million time steps).  $\sigma_0^2$ only influences the scale of the process and leaves the statistical properties of returns unchanged (a result confirmed in our simulations), so we do not show the results of varying $\sigma_0^2$ here.

In Fig.~1, we show the properties of the model with different $B$.  We set $\sigma_0^2=10^{-4}$ and let $B=10,B=100,B=1000$ respectively.  To facilitate the presentation of results, we define the following normalized variables:
\begin{eqnarray}
r'_t(\Delta t) & \equiv & (r_t(\Delta t) - \mu) / (\sigma_0 \sqrt{\Delta t}), \\
\sigma'_t & \equiv & \sigma_t/\sigma_0, \\
\beta'_t & \equiv & \sigma^2_0/\sigma^2_t.
\end{eqnarray}

\begin{figure*}
\includegraphics[width=3.4in]{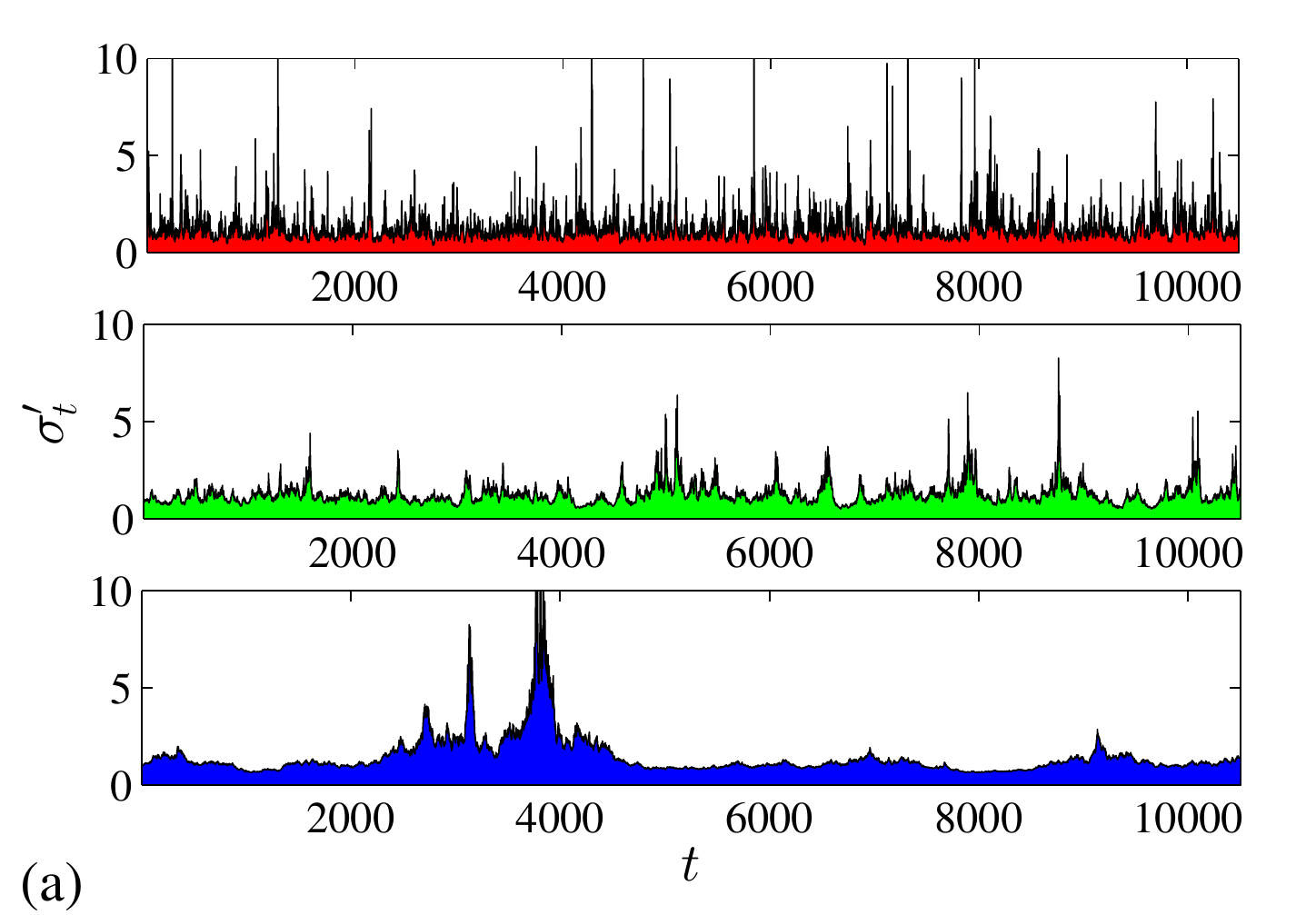}
\includegraphics[width=3.4in]{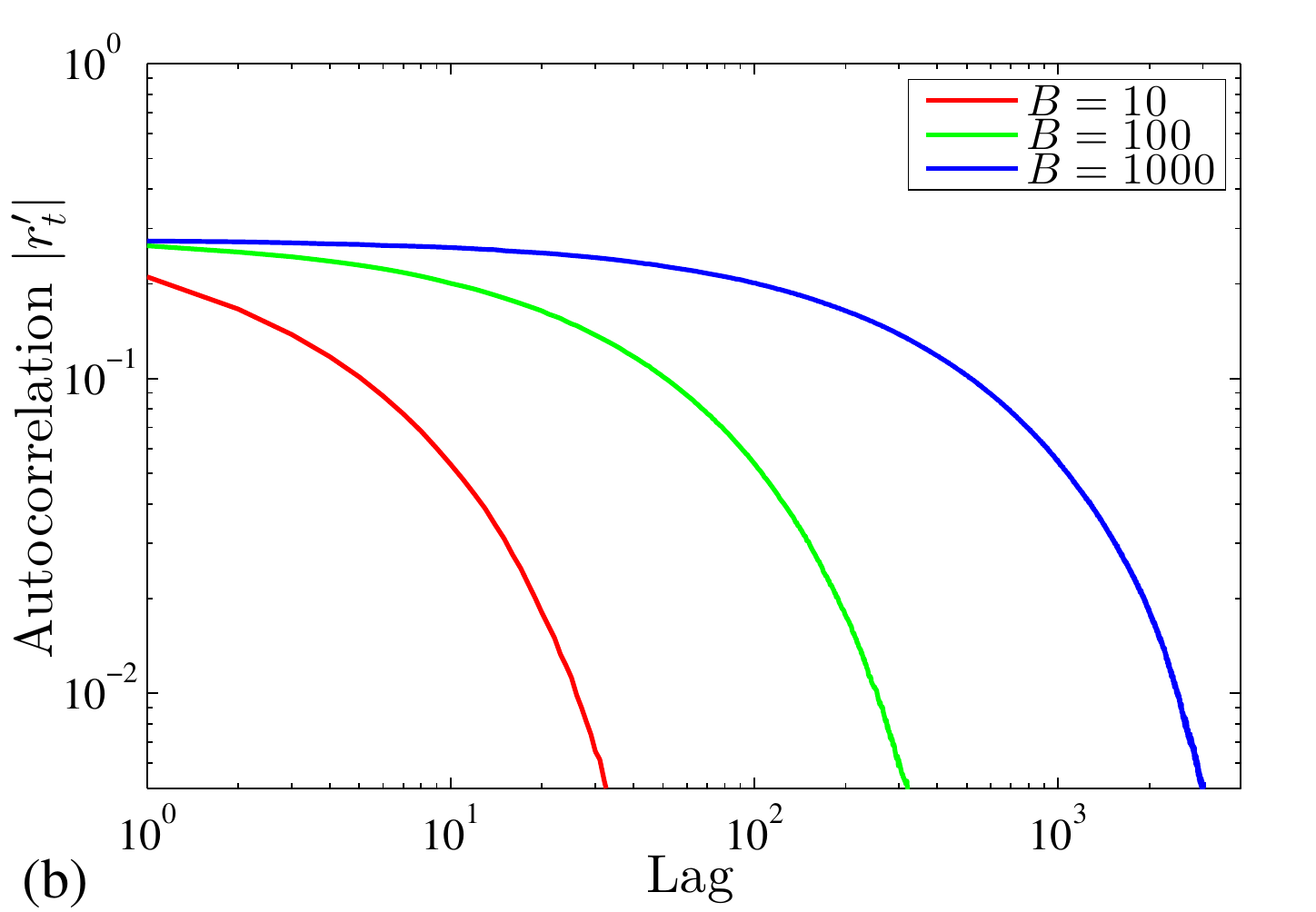}
\includegraphics[width=3.4in]{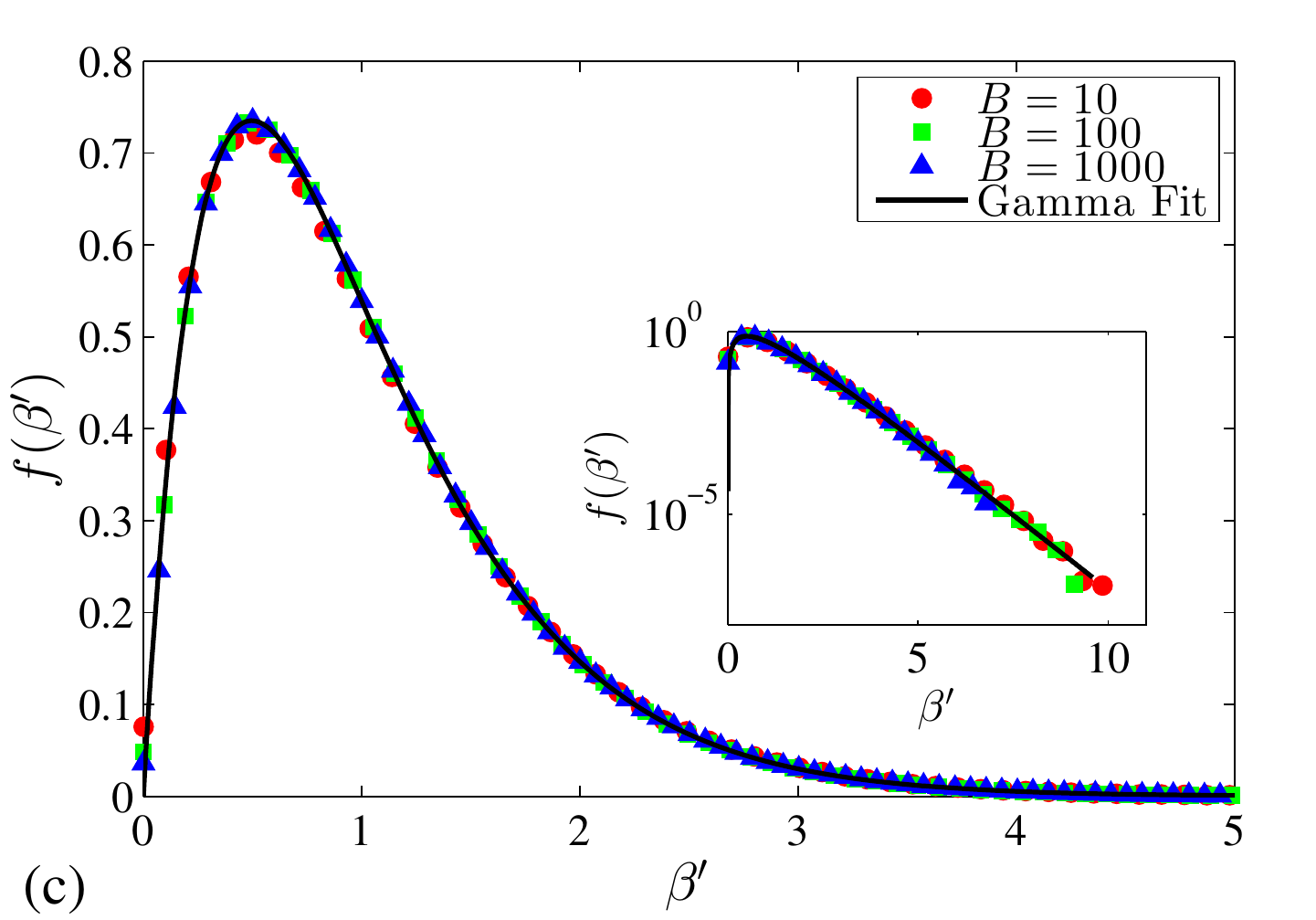}
\includegraphics[width=3.4in]{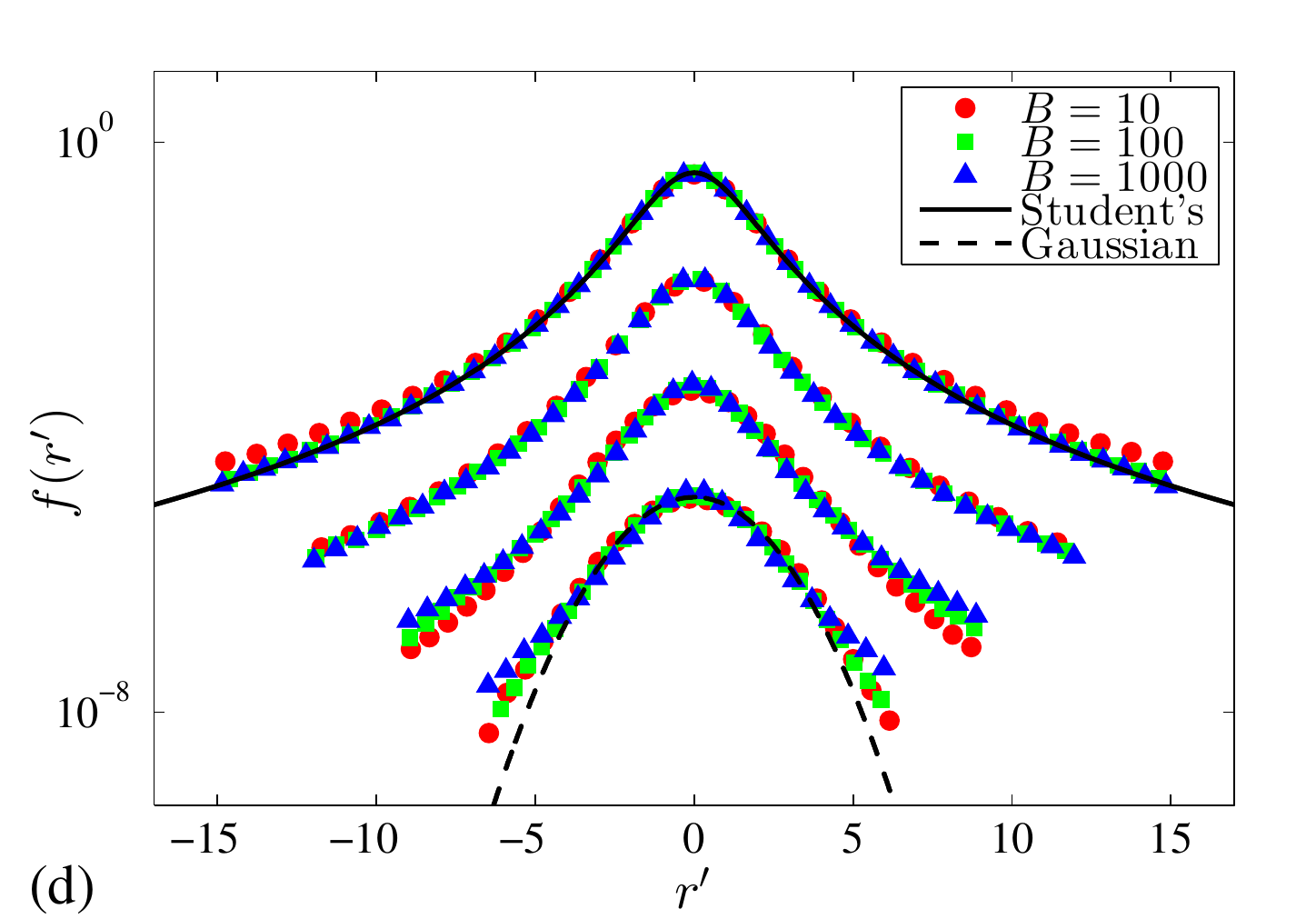}
\caption{Properties of the model with various choices of the parameter $B$. (a) The volatility series for $B=10$ (top), $B=100$ (middle), and $B=1000$ (bottom).  (b) The autocorrelation function of absolute scaled returns. (c) The probability density function of the scaled inverse variance (the inset plot is in semilog coordinates).  A fit to the Gamma distribution is shown for B=100. (d) The probability density function of scaled returns compared to a Student's $t$-distribution and a Gaussian.  The curves at different $\Delta t$ are arbitrarily offset vertically and from top to bottom are for $\Delta t = 1$,  $\Delta t = 10$, $\Delta t = 100$, and $\Delta t = 1000$.}
\end{figure*}

\begin{figure*}
\includegraphics[width=3.4in]{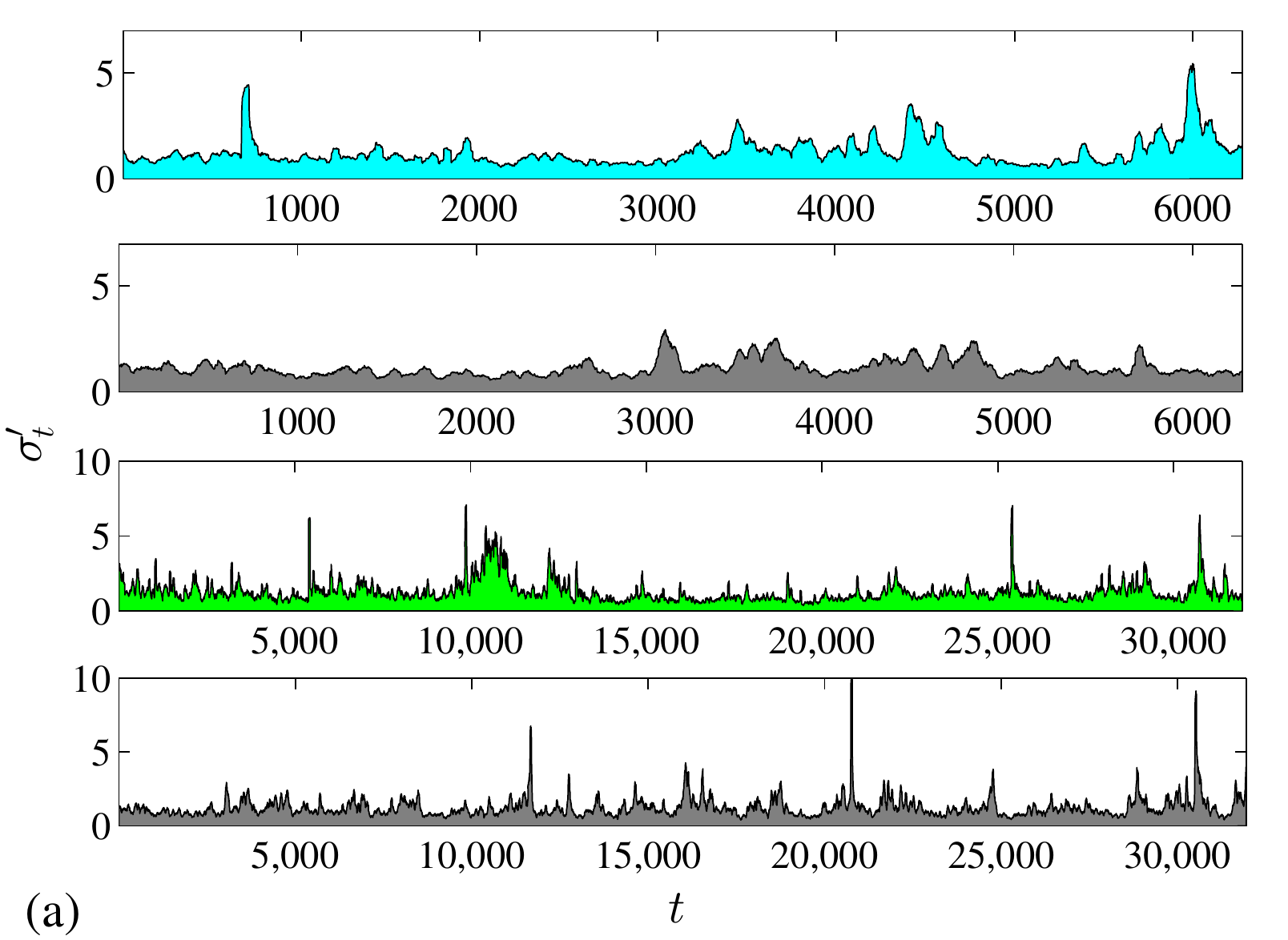}
\includegraphics[width=3.4in]{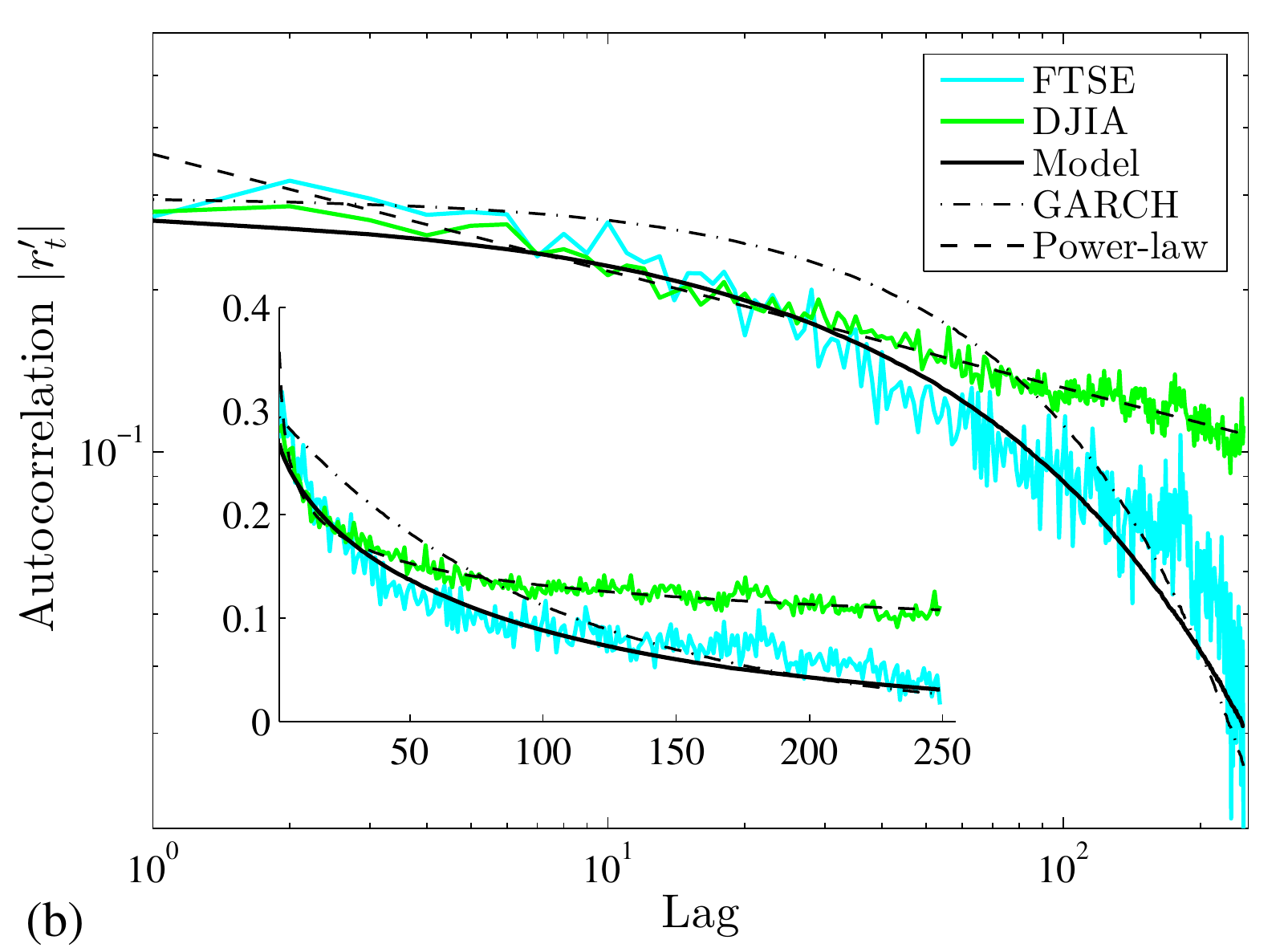}
\includegraphics[width=3.4in]{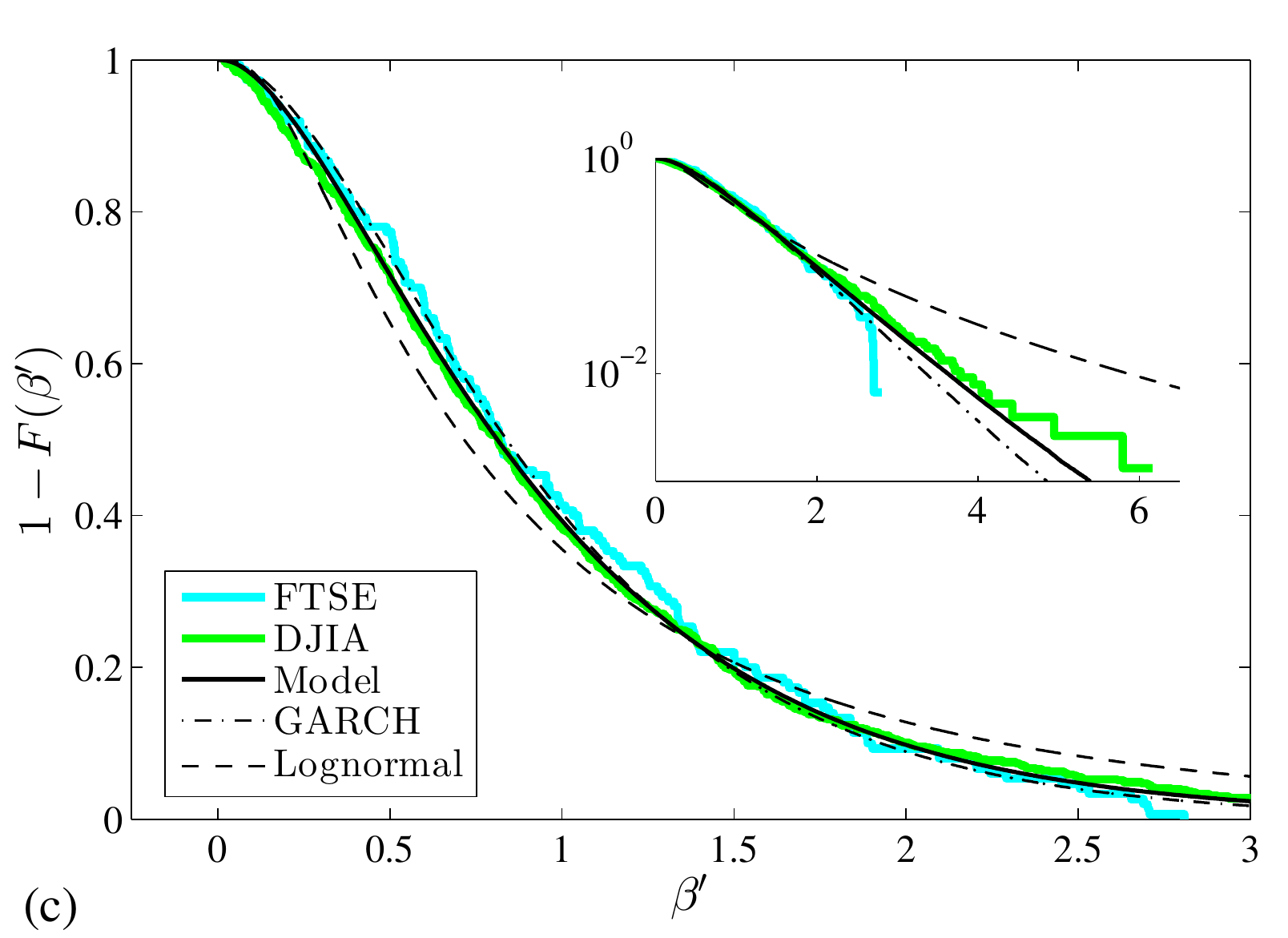}
\includegraphics[width=3.4in]{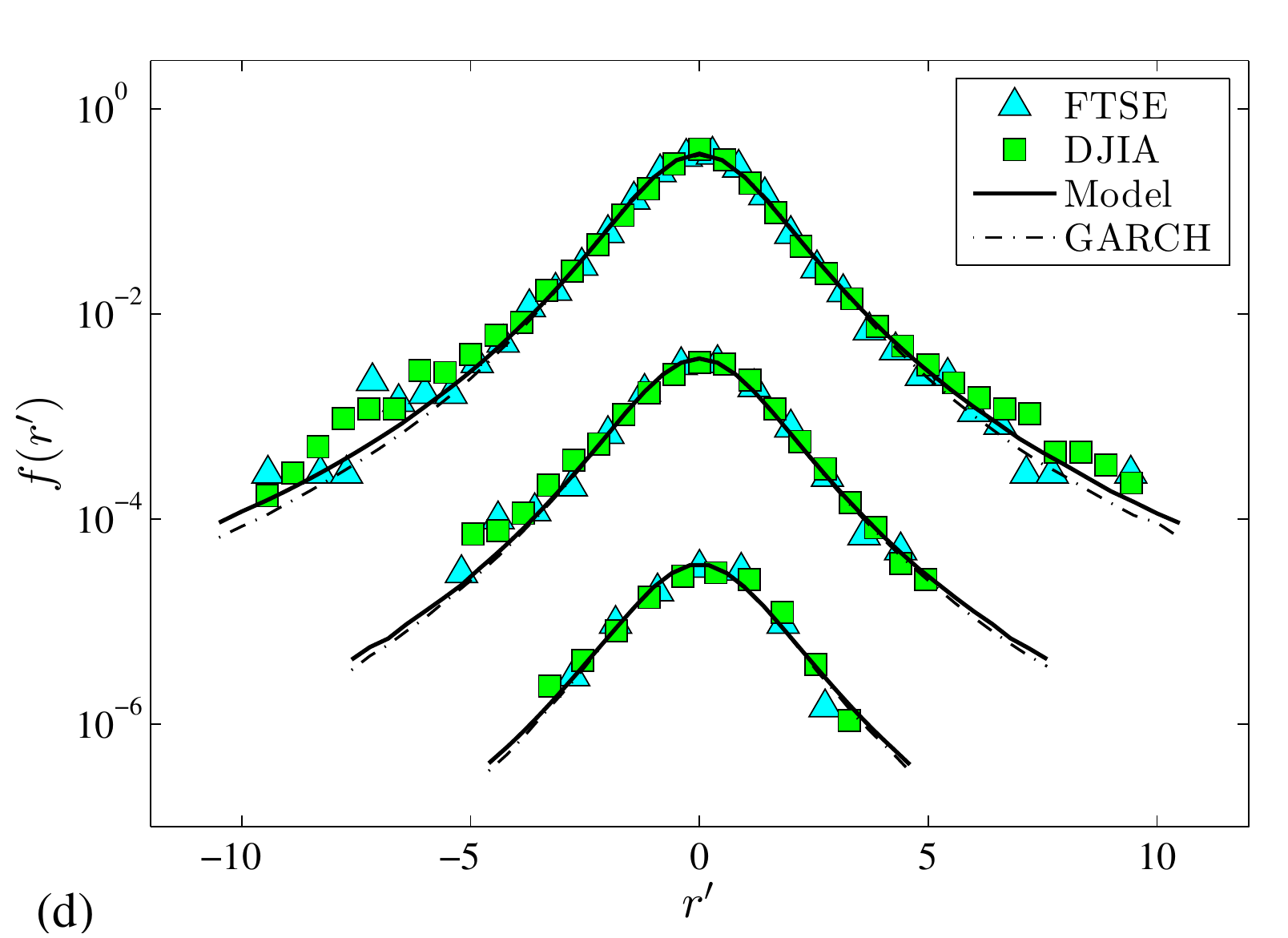}
\caption{Comparison of the model ($\sigma_0^2=1$, $B=164$) with data. (a) The volatility series for the FTSE data (top), the model over the same length of time as the FTSE data (top middle), the DJIA data (bottom middle), and the model over the same length of time as the DJIA data (bottom).  (b) The autocorrelation function of absolute scaled returns. The predicted autocorrelation function from a GARCH(1,1) process and a power-law fit both calibrated to the DJIA data are shown for comparison.  The inset plot is in regular coordinates. (c) 1 minus the cumulative distribution function of the scaled inverse variance.  The inset plot is in semilog coordinates.  The predicted distributions from a GARCH(1,1) process and a lognormal fit both calibrated to the DJIA data are shown for comparison. (d) The probability density function of scaled returns.  The curves at different $\Delta t$ are arbitrarily offset vertically and from top to bottom are for $\Delta t = 1$,  $\Delta t = 5$ (one week), and $\Delta t = 42$ (two months).}
\end{figure*}

As seen in Figs.~1(a and b), the parameter $B$ sets the strength of volatility autocorrelations with larger values of $B$ producing stronger autocorrelations.  As seen in Fig.~1(c), the distribution of the normalized inverse squared volatility, $\beta'$, is gamma distributed and largely unaffected by $B$ (although for $B=10$, the peak of the distribution is slightly below the others).   A gamma distributed $\beta'$ should produce $t$-distributed returns for $\Delta t=1$, which is observed in the topmost curves in Fig.~1(d).  As $\Delta t$ increases (moving to the lower curves in Fig.~1(d)), the return distribution adjusts from a Student's $t$-distribution to a Gaussian, with the speed of adjustment determined by $B$.  Higher values of $B$ correspond to a slowly varying volatility and therefore to a return distribution that retains its non-Gaussian shape at larger $\Delta t$.   

In Fig.~2 we compare the results of the model ($\sigma_0^2=1$, $B=164$) with two empirical financial time series: the daily price series of (a) the Dow Jones Industrial Average (DJIA) from May 26, 1896 to September 13, 2013 and (b) the FTSE 100 from January 2, 1985 to December 31, 2009.  The DJIA data is from the St. Louis Federal Reserve website and the FTSE data is from finance.yahoo.com.  From the price series, daily returns are calculated as in Eq.~1.  Squared volatilities are estimated using a rolling window of two months of daily returns (42 trading days),
\begin{equation}
\hat{\sigma}^2_t = \frac{\sum_{i=-21}^{20} (r_{t+i} - \hat{\mu})^2}{42},
\end{equation}
where $\hat{\mu}$ is the estimated mean of the daily returns.  We have systematically estimated the variance using window sizes from 10 to 100 days for the DJIA dataset.  The variance distribution changes as the window size increases, but reaches a steady distribution for sizes larger than 40.

We have found that a good estimate of $\sigma_0^2$, is the inverse of the mean inverse variance,
\begin{equation}
\hat{\sigma}_0^2 = \frac{1}{\left\langle 1 / \sigma_t^2 \right\rangle }.
\end{equation}
Using this equation, we find $\hat{\sigma}_0^2=5.1\times10^{-5}$ for the DJIA data and $\hat{\sigma}_0^2=6.3\times10^{-5}$ for the FTSE data.

We estimate $B$ for each series by minimizing the sum of the squared difference between $1/\sigma^2_t$ and its expected value, i.e., we find the $B$ that minimizes $\sum{e_t^2}$, where $e_t = 1/\sigma^2_t - (1+B\hat{\sigma}_0^2/\sigma_{t-1}^2)/((1+B)\hat{\sigma}_0^2)$.  Using this method, we find $\widehat{B}=164$ for the DJIA data and $\widehat{B}=167$ for the FTSE data.  

Although we obtain a different $\widehat{B}$ for the two empirical price series, the values are sufficiently similar that we choose to compare both datasets to the model using the DJIA fit, $B=164$ (see Fig.~2).  As in Fig.~1, the plots in Fig.~2 use the renormalizations of Eqs.~9-11 but with the estimated values of the parameters.  

To demonstrate how well the model matches data, we compare its predicted results to alternatives.  In Fig.~2(b) we compare the autocorrelation function from the model to that of a GARCH(1,1) process and also a power-law fit both calibrated to the DJIA data.  The model reproduces the autocorrelation function for both datasets until about 25 lags.  For higher lags, it continues to match the FTSE data but for the DJIA it is outperformed by the power-law fit.  Care should be taken when drawing conclusions from these results, however, as autocorrelation functions fluctuate considerably and are misleading if the data contains structural breaks\cite{Granger2004}.  Note that the model does not produce long-memory in volatility, i.e., the autocorrelation function of the model is integrable. 

For the distribution of the normalized inverse variance, we compare the model's predictions to that of a GARCH(1,1) process calibrated to each dataset (see caption of Table I for parameters) and also to a gamma, an inverse-gamma (the Heston model\cite{Heston1993} predicts a gamma distributed variance, i.e., an inverse-gamma distributed inverse variance), and a lognormal fit (which is often assumed in the finance literature: see \cite{Andersen01} and note that the inverse square of a lognormally distributed variable is also lognormally distributed).  We specifically show the GARCH(1,1) prediction for the DJIA data and the lognormal fit for the DJIA data in Fig.~2(c), and we test all of the predictions and fits using a chi-square goodness-of-fit test (see Table I).  In agreement with previous results\cite{Gerig2009,Fuentes2009}, the gamma distribution cannot be rejected for either dataset.  The predicted distributions from the model also cannot be rejected for either dataset (the predictions are also not rejected for the unnormalized distributions).  The distribution predicted by the GARCH(1,1) process is rejected for the DJIA data but not for the FTSE data (although not shown, the predictions for the unnormalized distributions are both rejected).  Both the lognormal and inverse gamma fits are rejected.

\begin{table}[htb]
\centering
Chi-square p-values
\begin{tabular}{lcccccccccc}
\hline
\hline
				&& Gamma 	&& Model	&& GARCH	&& Logn && InvGam	\\
\hline
FTSE		&&0.49		&&0.19		&&0.13\:\;\, \			&& 0.03*\, \		&& 0.00**\\
DJIA		&&0.42		&&0.38 		&&0.00**			&& 0.00** 	&& 0.00**\\
\hline
\end{tabular}
\caption{p-values of chi-square goodness-of-fit test for the normalized inverse variance distribution $f(\beta')$.  Starred entries represent rejection at the 5\% level and double starred entries at the 1\% level.  Parameters \{$a$, $b$\} of the gamma fits were \{2.0, 0.50\} and \{1.6, 0.6\} for the FTSE and DJIA data respectively.  Parameters \{$\mu$, $\sigma$\} of the lognormal fits were \{-0.27, 0.83\} and \{-0.33, 0.90\}.  Parameters \{$a$, $b$\} of the inverse gamma fits were \{1.3, 0.64\} and \{1.1, 0.45\}.  Parameters \{constant, $\sigma_{t-1}^2$ coeff., $r_{t-1}^2$ coeff.\} of the GARCH(1,1) process were \{1.5e-6, 0.90, 0.091\} and \{1.3e-6, 0.90, 0.089\}.}
\label{table.one}
\end{table} 

Financial time series have been studied by mathematicians and physicists over many years\cite{Bachelier64, Mandelbrot63, Mantegna99, BouchaudPotters03}.  Although the dynamics of prices are now well-characterized and understood, it is still unclear why prices exhibit the interesting properties that they do.  This lack of understanding is especially troublesome because prices fluctuate in a universal, regular way, i.e., the returns of many different traded items all possess the same non-trival properties.  It is therefore quite likely that some simple, robust mechanism underlies price dynamics, even if we have not yet discovered it\cite{Gerig2011}.

We have presented a simple feedback model for volatility that matches empirical data very well using only two parameters.  The model is unlike any other volatility model because our feedback mechanism is embedded in a parameter of the noise term, which is inverse gamma distributed.  Although we do not show it here, the model can be connected to the stochastic volatility literature because under certain restrictions, it reduces to the 3/2 stochastic volatility model\cite{PlatenHeath2006}.  Although also not shown, we have found that the model replicates the multifractal structure of returns and can be used to predict volatility better than the standard GARCH(1,1) model.  The model does not produce the well-known correlation between negative returns and volatility (known as the leverage effect) nor does it explicitly include feedback effects on multiple timescales\cite{Muller1997, LeBaron2001, Zumbach2001, Borland2011}, but these features could be added without difficulty.   

\begin{acknowledgments}
This work was supported by the European Commission FP7 FET-Open Project FOC-II (no. 255987) and a UTS Faculty of Business Research Grant.
\end{acknowledgments}

\end{document}